\documentclass[12pt]{article}

\begin{document}

\thispagestyle{plain}         
\setcounter{page}{1}         

\input amssym.tex

\newcommand{\Lor}{L^{\uparrow}_{+}}
\newsymbol\tss 1073

\newtheorem{lem}{Lemma}
\newtheorem{defin}{Definition}
\newtheorem{theor}{Theorem}
\newtheorem{rem}{Remark}
\newtheorem{prop}{Proposition}
\newtheorem{cor}{Corollary}
\newenvironment{demo}
{\bgroup\par\smallskip\noindent{\it Proof: }}{\rule{0.5em}{0.5em}
\egroup}

\title{Generalized Dirac monopoles in non-Abelian Kaluza-Klein theories}

\author{Ion I.  Cot\u aescu \thanks{E-mail:~~~cota@physics. uvt. ro}\\ 
{\small \it West University of Timi\c soara,}\\
       {\small \it V.  P\^ arvan Ave.  4, RO-300223 Timi\c soara, Romania}}

\date{}

\maketitle

\begin{abstract}

A method is proposed for generalizing the Euclidean Taub-NUT space, regarded as 
the appropriate background of the Dirac magnetic monopole, to non-Abelian 
Kaluza-Klein theories involving potentials of generalized monopoles.
Usual geometrical methods combined with a recent theory of the induced
representations governing the Taub-NUT isometries lead to a general
conjecture where the potentials of the generalized monopoles of any
dimensions can be defined in the base manifolds of suitable principal fiber
bundles. Moreover, in this way one finds that apart from the monopole models 
which are of a space-like type, there exists a new type of time-like models
that can not be interpreted as monopoles. The space-like
models are studied pointing out that the monopole fields strength are
particular solutions the Yang-Mills equations with central symmetry   
producing the standard flux of $4\pi$ through the two-dimensional spheres
surrounding the monopole. Examples are given of manifolds with Einstein
metrics carrying  $SU(2)$  monopoles.

Pacs 04. 62. +v

\end{abstract}

\newpage

\section{Introduction}

A special class of solutions of the Maxwell or Yang-Mills equations are the 
instantons and monopoles defined on  appropriate flat or curved backgrounds
\cite{AT,EGH}. Successful methods were used for investigating the geometric
properties of the original Dirac magnetic monopole \cite{D} and its Abelian
\cite{ABE} or non-Abelian generalizations \cite{YA,TP,MIN}. Moreover, the
study of the role of the BPS monopoles \cite{BPS} in the Lagrangian theories
with Higgs mechanisms is actually of a large interest \cite{BPS2}.

Another framework is offered by the Kaluza-Klein theories
where the Maxwell or Yang-Mills degrees of freedom deal with specific
extra-coordinates exceeding the physical spacetime. In these
theories the basic problem is to find solutions of the Yang-Mills equations
in a geometry whose global metric should be an exact solution of the
Einstein equations \cite{YME}.
A typical example is the four-dimensional Euclidean Taub-NUT space which
involves the potentials of the Dirac magnetic monopole and
satisfies the vacuum Einstein equations. Moreover, this geometry is
hiper-K\" ahler having many interesting properties related to a specific 
hidden symmetry \cite{NOVA}. For this reason the K\" ahlerian geometries were
considered for generalizing the Dirac monopole to many dimensions \cite{ABE}.        

In this article we should like to continue the  investigation of the
geometrical methods that could lead to new versions of generalized Dirac
monopoles in non-Abelian Kaluza-Klein theories of any dimensions. In this
matter there are different opinions concerning the possible topologies of
the backgrounds carrying generalized monopoles \cite{YY,BPS2}. Here we start
with the idea that the non-Abelian monopoles may have similar topologies and
the same type of invariants as the Abelian Dirac one. We assume that the
monopole field is (I) a particular solution of the Yang-Mills equations with
an apparent {\em string} singularity that (II) can be reduced up to a
point-like singularity grace to a suitable topology \cite{EGH}. This
remaining
singularity represents the monopole which must give rise to a field strength
with (III) a {\em central} symmetry up to gauge transformations and (IV)
producing through two-dimensional spheres surrounding the monopole the
standard flux of $4\pi$, in units of the coupling constant \cite{YA}. In
order to accomplish these requirements we shall focus on the relation between
the geometrical and the gauge symmetries \cite{SYM,YME}.

In a recent study concerning the isometries of the  Euclidean Taub-NUT space
\cite{IndR} we observed that the angular coordinates of the four-dimensional
Taub-NUT space form a set of parameters of the isometry group of the physical
three-dimensional subspace.  This property explains why the isometries of the
Euclidean Taub-NUT space involve linear rotations of the
Cartesian physical coordinates while the  extra-coordinate transforms
according to a non-linear representation of the rotation group. We have found
the integral form of this representation showing that this is
induced by the subgroup of the rotations that preserve the direction of
the string giving rise to the Dirac monopole. 

Here we intend to generalize this conjecture combining our method of induced
representations \cite{IndR} with the usual theory of the principal fiber
bundles. We consider that the manifolds of our non-Abelian Kaluza-Klein
theories are principal bundles where the fibers are orthogonal groups and
the base manifolds are physical spaces with a manifest central symmetry.
In this context we propose the concrete integral form of the isometry
transformations and  derive the representations induced by the structure
group of the principal bundle in the general case of any
(pseudo)-orthogonal isometry group. Hereby we deduce the gauge
transformations associated to isometries that guarantee the central symmetry
of the entire theory and offer us the tools for eliminating 
the original string singularity within a suitable non-trivial fibration
\cite{EGH,MIN,HOPF}. The
remaining monopole singularity gives rise to fields strength  with central
symmetry that satisfy the sourceless Yang-Mills equations. Moreover, we show
that these solutions are  topologically {\em stable} in the sense that 
their principal bundles are non-trivial only over a two-dimensional sphere
surrounding the monopole \cite{BPS2}. Then it is natural to find that the
principal invariant of these models is just the desired standard flux.

We shall achieve these general objectives, starting in the second section
with a short presentation of the lesson offered by the Euclidean Taub-NUT
geometry that will guide us to a natural generalization of the monopole 
geometry. The next section is devoted to some technical details 
concerning the method of introducing coordinates in orbits related to the 
Lie groups and involved in the fiberings we use. In addition, the horizontal and 
vertical projections in the tangent spaces of the principal bundles are also 
considered. Our main results are obtained in the fourth section where we 
introduce the isometries of our approach which generate the gauge
transformations we need for giving up the effects of the string singularities
without to affect the global central symmetry. Moreover, we show that beside
the usual monopole models having the strings in a space-like direction, there
exists another type of models with time-like strings but correct unitary gauge
groups. However, since these models are rather unusual, we analyze in the
fifth section only the space-like models obtaining the general solutions
for the potentials and the corresponding fields strength. We show how can
be eliminated
the string singularity and study the topological stability of the monopole
calculating the flux of its field strength. Examples of $SU(2)$ monopoles in
concrete Einstein spaces are given before to present the final comments.
                 
We use the natural units with $\hbar=c=1$ and consider unit coupling
constants.

\section{The lesson of the Taub-NUT geometry}

The Euclidean Taub-NUT manifold is the space of the Abelian Kaluza-Klein
theory of the Dirac magnetic monopole that  provides a non-trivial 
generalization of the Kepler problem.  This space is a special member of the 
family of four-dimensional manifolds, ${ M}_4\sim {\Bbb R}^4$, equipped with 
the isometry group ${\bf I}({ M}_4)=SO(3)\otimes U(1)$ and carrying the 
the Dirac magnetic monopole related to a string along the third axis.  

These geometries can be easily constructed in local charts with spherical 
coordinates $(r,\theta, \varphi, {\alpha})$ among them the first three are the 
usual spherical coordinates of the vector $\vec{x}=(x^1,x^2,x^3)$, with 
$|\vec{x}|=r$, while ${\alpha}$ is the Kaluza-Klein extra-coordinate of this chart.  The spherical 
coordinates can be associated with the Cartesian ones $(x,y)=(x^1,x^2,x^3,y)$ 
where the extra-coordinate is up to a factor $y=-({\alpha}+\varphi)$.  
In Cartesian coordinates one has the opportunity to use the vector notation 
and the scalar products $\vec{x}\cdot \vec{x}^{\prime}$ which are invariant 
under the $SO(3)$ rotations.       

The group $SO(3)\subset {\bf I}({ M}_4)$ has three independent one-parameter 
subgroups, $SO_i(2)$, $i=1,2,3$, each one including rotations ${ R}_i(\psi)$, 
of angles $\psi\in [0,2\pi)$ around the axis $i$.  With this notation any 
rotation ${ R}\in SO(3)$ in the usual Euler parametrization reads 
${ R}(\alpha,\beta,\gamma)={ R}_3(\alpha){ R}_2(\beta){ R}_3(\gamma)$.
Moreover, we can write $\vec{x}={ R}(\varphi,\theta,0) \vec{x}_o$ where the  
vector $\vec{x}_o=(0,0,r)$ along the string direction is invariant under the 
rotations $R_3\in SO_3(2)$.   For this reason one says that $SO_3(2)$ is 
the little group of the string direction  $\vec{x}_o$.  This group will be the 
second term of the isometry group ${\bf I}(M_4)$ usually denoted by $U(1)$ 
since these two groups are isomorphic.  

We have shown \cite{IndR} that the transformations of the isometry group 
${\bf I}(M_4)$ can be written explicitly in an integral form defining the 
action of two arbitrary rotations, ${ R}\in SO(3)$ and 
${ R}_3\in SO_3(2)\sim U(1)$, in the spherical 
charts,  $({ R},{ R}_3) : (r, \theta, \varphi, {\alpha}) \to (r, \theta', \varphi', {\alpha}')$, 
such that 
${ R}(\varphi', \theta', {\alpha}')={ R}\,{ R}(\varphi,\theta,{\alpha})
{ R}_3^{-1}$. 
Hereby it results that the Cartesian coordinates transform under 
rotations ${ R}\in SO(3)$ as 
\begin{eqnarray}
\vec{x}&\to&\vec{x}'={ R}\,\vec{x}\,, \label{lin}\\
{y}&\to& {y}'={y}+{h}({ R},\vec{x})\,,\label{ind} 
\end{eqnarray}
where the function ${h}$ is given in Ref.  \cite{IndR}.  Thus, the vector 
$\vec{x}$ transforms according to an usual linear representation but 
the transformation of the fourth Cartesian coordinate is governed by a 
representation of $SO(3)$ induced by $SO_3(2)$.   
The transformations of the group $SO_3(2)\sim U(1)$ affect only the fourth 
coordinate translating it.

In this context, we observe that the 1-forms
\begin{equation}
d\Omega(\varphi, \theta, {\alpha})=
{ R}(\varphi, \theta, {\alpha})^{-1} d{ R}(\varphi,\theta,{\alpha}) \in so(2)
\end{equation} 
transform independently on ${ R}$ as
\begin{equation}
({ R},{ R}_3) : d\Omega(\varphi, \theta, {\alpha})
\to d\Omega(\varphi', \theta', {\alpha}')={ R}_3 d\Omega(\varphi,\theta,{\alpha})
{ R}_3^{-1}\,,
\end{equation}
finding that, beside the trivial quantity ${ds_1}^2=dr^2$, there are two 
types of line elements invariant under ${\bf I}({ M}_4)$,
\begin{eqnarray}
&&{ds_2}^2=-\left< d\Omega(\varphi,\theta,{\alpha})^2\right>_{33}=
d\theta^2+\sin^2\theta d\varphi^2\,,\\
&&{ds_3}^2=-\frac{1}{2}{\rm Tr}\left[ d\Omega(\varphi,\theta,{\alpha})^2\right]=
d\theta^2+\sin^2\theta d\varphi^2+(d{\alpha}+\cos\theta d\varphi)^2\,. \label{g3}
\end{eqnarray}
Here it is worth pointing out that the above metrics are related to a family 
of metrics on the spheres $S^3\sim SO(3)$ parametrized with the angle
variables
$(\theta,\varphi,{\alpha})$ with $ 0\leq\theta<\pi, 0\leq\varphi<2\pi$ and  
$0\leq{\alpha}<4\pi$.  The Hopf fibration $(S^3,{ds_3}) \to 
(S^2, {ds_2})$  defines a vertical subbundle  and its 
orthogonal complement  with respect to the standard metric 
(\ref{g3}).  The restrictions of this metric to the horizontal, 
respectively the vertical bundle, give  the horizontal  line element 
$d{s_h}^2={ds_2}^2$ and the vertical one 
\begin{equation}\label{dsV}
{ds_v}^2={ds_3}^2-{ds_2}^2 =(d{\alpha} + \cos\theta d\varphi)^2\,.       
\end{equation}

The conclusion is that the manifolds $(M_4, ds)$ with the isometry group 
${\bf I}(M_4)$  have metrics defined by the line elements 
of the general form, 
\begin{equation}\label{ds}
ds^2={ds_o}^2+  f_v(r){ds_v}^2\,,\quad 
{ds_o}^2=f(r)dr^2+  f_h(r){ds_h}^2\,, 
\end{equation}
where  $f$, $f_h$ and $f_v$ are arbitrary functions only on $r$.   
From the physical point of view these manifolds represent backgrounds of 
Abelian Kaluza-Klein theories where the physical spaces (without gauge fields)
are the base manifolds
of the fibrations $(M_4, ds)\to (M_3, {ds_o})$ induced by the Hopf one.  The 
fiber results to be isomorphic with the little group $SO_3(2)\sim U(1)$ which 
plays the role of the {\em gauge} group of these Abelian theories.  
The form of the vertical  metric (\ref{dsV}) in Cartesian coordinates, 
${ds_v}^2=(d{y} + A_i dx^i)^2$, emphasizes  of the potentials $A_i$  
of the Dirac magnetic monopole.  

Hereby the lesson we have to learn is that some  Kaluza-Klein theories could
be obtained {\em directly} constructing the corresponding geometries with an
appropriate symmetry.  The main point is that the physical angular
coordinates $(\theta, \varphi)$ and the Kaluza-Klein extra-coordinate
${\alpha}$ form a set
of Euler parameters of the $SO(3)$ group.  This allowed us to define the 
isometries in integral form and to find the general expression of the invariant 
line elements (\ref{ds}).  It is important to observe that the Hopf fibration,
involving only spheres, removes the string singularities of the potentials 
$A_i$ in all the geometries of this class of symmetry, independent 
on the choice of the invariant functions $f$, $f_h$ and $f_v$.    

This conjecture can be generalized to non-Abelian Kaluza-Klein theories with
$n$-dimensional physical spaces. We shall assume that in the whole manifolds
of these theories there exist  orbits $\Sigma \sim SO(n)$ whose coordinates,
including the extra-dimensions, form  sets of parameters of the groups
$SO(n)$. In any physical space, we have to take a string with the
corresponding little group $SO(n-1)$ that preserves its direction and defines  
the orbits $\Sigma_o\sim SO(n)/SO(n-1)$ of the physical space. Then
we can exploit the fibrations $\Sigma\to \Sigma_o$ whose fibers
$F\sim SO(n-1)$ are isomorphic with the little groups. In this framework the
isometries can be introduced postulating similar automorphysms as in the
previous particular case. Thus we may obtain $SO(n)\otimes SO(n-1)$
isometries and associated gauge transformations which should assure the
central symmetry of the whole theory and convenient non-trivial fiberings.
These geometries have to give rise to particular Yang-Mills potentials that
may be written down separating the vertical projections of the metrics in the
tangent spaces of the principal bundles $\Sigma$.
We note that in this approach we must consider, in addition, the associated
spin bundles whose structure groups are the universal covering groups
${\rm Spin}(n-1)$ of the structure groups $SO(n-1)$ of the principal bundles.

In what follows we should like to develop this theory for any dimensions and
pseudo-orthogonal groups.

\section{The general framework}

Let us start with the vector space $M_n\sim {\Bbb R}^n$ equipped with the 
{\em associated} (pseudo)-Euclidean metric $\eta$ of an arbitrary signature 
$(n_+, n_-)$ with $n_+ + n_-=n$ that for $n_-=0$ becomes Euclidean.  
We denote by 
$x^{\mu}$, $\alpha,. . . , \mu,\nu,. . . =1,2,. . . ,n$ the 
components of a vector $x=(x^1,x^2,. . . ,x^n)^T$ and consider the bilinear 
form $(~\cdot~):M_n\times M_n \to {\Bbb R}$ defined by $\eta$ 
that can be expressed  in the matrix notation as $(x\cdot x')=x^T \eta x'$.  
This form remains invariant under the transformations, 
$G:x\to Gx\,, \forall\, x\in M_n$, of the fundamental representation of 
the (pseudo)-orthogonal group ${\bf G}=SO(n_+ ,n_-)$ since the matrices 
$G\in {\bf G}$ obey $G^T\eta G=\eta$.  We assume that $M_n$ is orientable, the 
improper transformations of the group $O(n_+,n_-)$ changing the 
chirality of its Cartesian frames.  When $(x\cdot x)>0$ we can use  the 
unit vector $\hat x$ defined such that $x=|x|\hat x$ where  the invariant
$|x|=\sqrt{(x\cdot x)}$ becomes the norm of $x$ in the Euclidean case. 

\subsection{Groups and orbits}

In ${\bf G}$ one can introduce many types of parametrizations but in practice 
it is convenient to work with the standard covariant parametrization (of the 
first kind),  
\begin{equation}\label{GX}
G(\omega)=e^{X(\omega)}\,, \quad X(\omega)=\textstyle\frac{1}{2}
X_{(\alpha\beta)}\omega^{\alpha\beta}\,,\quad \forall\, G\in {\bf G}
\end{equation}
with real skew-symmetric parameters $\omega^{\mu\nu}=-\omega^{\nu\mu}$ 
and real basis-generators $X_{(\mu\nu)}=-X_{(\nu\mu)}$ whose properties are 
briefly presented in the Appendix A. 
The matrices $X(\omega)$ form the fundamental representation of the Lie 
algebra ${\bf g}=so(n_+,n_-)$ of the group ${\bf G}$ carried by the vector 
space $M_n$.  When one uses other parametrizations one can 
express $\omega=\omega(z)$ in terms of the new parameters $(z)$ and  
work with the generators $X(z)=X[\omega(z)]$ without to change the 
basis-generators of ${\bf g}$.  The matrices $X(z)$ have to generate the 
transformations matrices $G(z)=G[\omega(z)]$ according to Eq. (\ref{GX}).
However, other types of parametrizations similar to the Euler one will be
also used. In any parametrization  $G(0)=1_n$ is the identity matrix of
${\bf G}$.

For our further developments the 1-forms $d\Omega(G)=G^{-1} dG$ defined for 
any $G\in {\bf G}$ are of a major importance.  These are elements of the 
Lie algebra ${\bf g}$ that satisfy 
\begin{equation}\label{OO}
d\Omega(G'G)= d\Omega(G)+ G^{-1}d\Omega(G')G\,,\quad   
d\Omega(G^{-1})=-G\, d\Omega(G)G^{-1}   
\end{equation}
and have the supplemental property
\begin{equation}\label{OT}
d\Omega(G)^T=-\eta\, d\Omega(G)\eta\,,  
\end{equation}
resulting from the fact that $G$ are (pseudo)-orthogonal matrices.   

It is know that, in general, any Lie group ${\bf G}=SO(n_+,n_-)$ is isomorphic 
with a manifold $\Sigma$ of dimension $\frac{1}{2}n(n-1)$ which is an 
hyperboloid for an arbitrary signature $(n_+,n_-)$ or a sphere in the Euclidean 
case when $\eta=1_n$ and ${\bf G}=SO(n)$ is an orthogonal group.  This means 
that a given parametrization $(z)$ of ${\bf G}$ defines a system of coordinates in a local chart of 
$\Sigma\sim {\bf G}$.  The general expression of the line element in this chart 
is given by the Killing metric as
\begin{equation}\label{sZ}
{ds_{Z}}^2=-{\rm Tr}\{Z\,d\Omega[G(z)]^2\}\,,
\end{equation}
where $Z=Z^{\dagger}$ is a Hermitian non-singular matrix which plays the role 
of a metric operator.  In other respects, the manifolds $\Sigma$ can be seen as 
being embedded in a larger space ${ M}_N\sim {\Bbb R}\times \Sigma \sim 
{\Bbb R}^N$ of dimension $N=\frac{1}{2}n(n-1)+1$ where the set of shells 
$\Sigma$ of different radius  may represent a foliation of a several domain of 
${ M}_N$ or even of the whole space ${ M}_N$ when the metric $\eta$ is 
Euclidean and  $\Sigma\sim S^{N-1}$ are spheres.  

Similarly, we  consider the orbits of the group ${\bf G}$ in the carrier space 
$M_n$.  For the {\em fixed} vector $x_o$ with $(x_o\cdot x_o)>0$ there exists 
the associated orbit $\Sigma_o=\{x|x=Gx_o,\, \forall\, 
G\in {\bf G}\}\subset M_n$ formed by all the vectors $x$ with the 
same norm $|x|=|x_o|$.  The 
transformations $G_o$ that leave $x_o$ invariant constitute the little group 
${\bf G}_o=\{G_o|G_o x_o =x_o\}$ of the orbit $\Sigma_o$ which will be 
supposed to be a {\em compact} subgroup of ${\bf G}$. 
All the orbits $\Sigma_o$ of different radius  are  isomorphic with the
coset space ${\bf G}/{\bf G}_o$ and, therefore, ${\rm dim}\,\Sigma_o=n-1$.   
In general, these are hyperboloids but when the metric is Euclidean and 
${\bf G}$ is compact then the orbit $\Sigma_o \sim S^{n-1}$ is a  sphere of 
radius $|x_o|$ in $M_n$.  In this case $\Sigma_o$ is covered by two local charts, 
one for the upper hemisphere $\Sigma_o^{+}$ with the pole in $x_o$ 
and another for the lower hemisphere $\Sigma_o^{-}$ whose pole is in 
$-x_o$.   We say that the $n-1$ parameters defining the unit vector $\hat x$ 
form the Cartesian coordinates $(\hat x)$  of the orbit $\Sigma_o$ 
understanding that for the spherical orbits we use the same coordinates in the 
domain $\Sigma_o^{+}\bigcap \Sigma_o^{-}$.  
  
We have seen that the  central pieces of the geometries we  study here are the 
string along the direction of $\hat x_o$ and the corresponding little group
${\bf G}_o$.  This suggests us to consider the well-known fibration
${\bf G}\to {\bf G}/{\bf G}_o$ that corresponds to the fibration $\Sigma
\to \Sigma_o$ the base of which is the orbit $\Sigma_o$ of the space $M_n$.
In this way the orbit $\Sigma$ becomes a principal fiber bundle with the
fiber  $F\sim {\bf G}_o$. This fibering  splits the tangent space
${\cal T}(\Sigma)\sim {\bf g}$ of $\Sigma$ into a vertical subbundle and its
horizontal complement. In terms of Lie algebras, the vertical projection is
the Lie algebra ${\bf g}_o$ of ${\bf G}_o$ which is a subalgebra in ${\bf g}$ 
while the horizontal complement is the coset space ${\bf g}/{\bf g}_o$. 

These may be easily manipulated if we introduce the elementary projection
operator $P_o$ on the one-dimensional subspace of $x_o$, with the
obvious properties $P_o x_o =x_o$ and $P_o G_o=G_o P_o=P_o$ for any 
$G_o\in {\bf G}_o$.  Since $P_o$ is an elementary projection 
operator it gives the signature $\epsilon_o$ of the direction 
$\hat x_o$ as 
\begin{equation}\label{Poeta}
P_o\eta = \eta P_o=\epsilon_o P_o.      
\end{equation}
The operator $P_o$ allows us to separate the elements of the Lie algebra 
${\bf g}_o$.  Thus any generator $X\in {\bf g}$ admits the orthogonal
decomposition, $X=X_v+X_h$, in its vertical and horizontal parts expressed as    
\begin{eqnarray}                
X_v&=&(1_n-P_o)X(1_n-P_o)=X-P_oX-X P_o \in {\bf g}_o\,,\\
X_h&=&P_oX(1_n-P_o)+(1_n-P_o)XP_o=P_oX+XP_o \in {\bf g}/{\bf g}_o\,. \label{XHini}
\end{eqnarray}
Here we used the property that $P_oXP_o=0$ for all $X\in {\bf g}$ since
$P_o$ is an elementary projection operator and $X$ has no diagonal elements. 
Other useful calculation formulas are
\begin{equation}\label{TrXH}
{\rm Tr}({X_h}^2)=2{\rm Tr}(P_o X^2)\,, \quad 
{\rm Tr}({X_v}^2)={\rm Tr}(I_o X^2)\,,
\end{equation}   
where  $I_o=1_n-2P_o$ is the reflexion transformation of the group
$O(n_+,n_-)$ which changes the sign of the string axis, $I_o x_o=-x_o$. 

\subsection{Cartesian coordinates}

The orbit $\Sigma_o$ may be generated either using arbitrary transformations
of ${\bf G}$ or explicitly exploiting a {\em fixed} isomorphism 
$B : \Sigma_o\to {\bf G}/{\bf G}_o$. This gives the "boosts" matrices 
$B(\hat x)$, that transform $x_o$ 
into a desired vector $x=B(\hat x)x_o$.  Since $|x|=|x_o|$ the matrix $B(\hat x)$ 
depends only on the $n-1$ parameters of $\hat x$ which are just the coordinates 
of the Cartesian chart $(\hat x)$ of the orbit $\Sigma_o$.  We assume that
the boost matrices have the form
\begin{equation}\label{Bx}
B(\hat x)=e^{\lambda\,X(\hat x)}\in {\bf G}/{\bf G}_o\,,
\end{equation}
depending on the scalar function  $\lambda=\lambda(\hat x)$ and the 
{\em horizontal} generator defined by 
\begin{equation}\label{XH}
X(\hat x)=\frac{X_{(\alpha \beta)}\hat x^{\alpha}\hat x_o^{\beta}}
{\sqrt{|1-(\hat x\cdot \hat x_o)^2|}} \in {\bf g}/{\bf g}_o\,,
\end{equation}
for all $\hat x \not=\hat x_o$ whereas $X(\hat x_o)=0$.  
Their properties given in the Appendix A allow us to find the closed matrix
form
\begin{equation}
B(\hat x)=\left\{
\begin{array}{lll}
1_n+X(\hat x)^2(1-\cos \lambda)+X(\hat x)
\sin\lambda&{\rm if}&(\hat x\cdot\hat x_o)<1\,,\\
1_n+X(\hat x)^2(\cosh \lambda-1)+X(\hat x)
\sinh\lambda&{\rm if}&(\hat x\cdot\hat x_o)>1\,,
\end{array}\right. 
\end{equation}      
where in the first case we identify $\cos \lambda=(\hat x\cdot \hat x_o)$ 
while in the second one  we must take $\cosh \lambda=
(\hat x\cdot \hat x_o)$.   In other words $B(\hat x)$ can be either a 
rotation of angle $\lambda$ or a Lorentz-type boost.  
These boosts will help us to calculate the line elements of the charts with 
Cartesian coordinates  $(\hat x)$ of $\Sigma_o$ in terms of the 1-forms
\begin{equation}\label{split} 
d\Omega[B(\hat x)]= d\Omega[B(\hat x)]_h +d\Omega[B(\hat x)]_v \,,
\end{equation}
whose projections are
\begin{equation}
d\Omega[B(\hat x)]_h=\left\{
\begin{array}{lll}
d\lambda X(\hat x)+\sin \lambda\, dX(\hat x)
&{\rm if}& (\hat x\cdot \hat x_o)<1\,,\\
d\lambda X(\hat x)+\sinh \lambda\, dX(\hat x)
&{\rm if}& (\hat x\cdot \hat x_o)>1\,,\\
\end{array}\right. \quad
\end{equation}
and
\begin{equation}\label{dOV}
d\Omega[B(\hat x)]_v=\left\{
\begin{array}{lll}
(1-\cos\lambda)[d X(\hat x)\,, X(\hat x)]
&{\rm if}&(\hat x\cdot\hat x_o)<1\,,\\
(\cosh\lambda-1)[d X(\hat x)\,, X(\hat x)]
&{\rm if}&(\hat x\cdot\hat x_o)>1\,.
\end{array}\right. 
\end{equation}      
We note that these formulas were obtained taking into account that for any 
generator $X =X(\hat x)$ we have $X\,dX\,X = (\hat x \cdot d\hat x)\, X=0$.    
Then using the general properties (\ref{OO}), (\ref{OT}) and (\ref{Poeta}) we 
find the invariant line element 
\begin{eqnarray}\label{dsHo}
{ds_h}^2&=&(d\hat x\cdot d\hat x)= -\hat x_o^T\eta d\Omega[B(\hat x)]^2 \hat x_o
\nonumber \\
&=& -\epsilon_o {\rm Tr}\left\{P_o d\Omega[B(\hat x)]^2\right\}
= -\frac{\epsilon_o}{2} {\rm Tr}\left\{d\Omega[{B(\hat x)]_h}^2\right\}\,. 
\label{dsH}
\end{eqnarray}

The Cartesian coordinates of the orbit $\Sigma$ will be defined bearing in 
mind that $\Sigma$ is isomorphic with ${\bf G}$.  Therefore,  we introduce the 
Cartesian coordinates $(\hat x, y)$ in a local chart of $\Sigma$ using the 
following parametrization of the group ${\bf G}$
\begin{equation}\label{Cart}
G(\hat x, y)=B(\hat x)G_o(y) \,,
\end{equation}
where $(y)$ represents a set of $m=\frac{1}{2}(n-1)(n-2)$ arbitrary parameters
of the group ${\bf G}_o$.  In these coordinates  the metric of $\Sigma$ 
is given by a line element of the general form (\ref{sZ}) where we must take 
$(z)\equiv (\hat x,y)$.  Then, we say that Eq. (\ref{dsHo}) defines the 
horizontal projection of this metric.

We have introduced the parametrization (\ref{Cart}) in accordance to the
fibration $\Sigma \to \Sigma_o$ with the base $\Sigma_o$.
Whenever the group ${\bf G}$ is not compact the orbit
$\Sigma_o$ may have different disjoint shells each one being covered by its
own chart and, therefore, the fibration is {\em trivial}.  In the compact case the
situation is more complicated since then the orbit $\Sigma_o$ is a sphere
with at least two local charts where we can take different parameters of the
fiber $F$.  Therefore, we introduce the Cartesian coordinates 
$(\hat x,y^{+})$ for the upper hemisphere $\Sigma_o^+$ and $(\hat x, y^{-})$ 
for the lower one, $\Sigma_o^-$.  On the overlapping domain 
$\Sigma_o^+\bigcap \Sigma_o^-$ these coordinates must be related among 
themselves through the  transition 
\begin{equation}\label{GHG}
G_o(y^-)={T}(\hat x)G_o(y^+)\,, \quad   
{T}(\hat x)\in  {\bf G}_o\,.
\end{equation}
The section $T:\Sigma_o \to {\bf G}_o$ is the transition function \cite{EGH}
of the fibration $\Sigma \to \Sigma_o$ which determines the topological
properties of the principal bundle $\Sigma$, including its homotopy type.
   
\subsection{Spherical coordinates}

Other important systems of coordinates of $\Sigma$ involved in applications 
are the spherical coordinates. These have to be introduced observing that the 
above boosts can be written as
\begin{equation}\label{BGBG}
B(\hat x)=G_o(\theta)B_s(\lambda)G_o(\theta)^{-1} \,,
\end{equation}  
where $B_s(\lambda)$ is a one-parameter transformation matrix of the form 
(\ref{Bx}) generated by the fixed horizontal generator $X_s\in {\bf g}/{\bf g}_o$ 
and arbitrary $\lambda$.  This transforms  $\hat x_o$ into the unit vector  
$\hat x(\lambda)=B_s(\lambda)\hat x_o$ while $G_o(\theta)\in {\bf G}_o$ 
performs the rotation $\hat x=G_o(\theta)\hat x(\lambda)$.  There are $n-2$ 
parameters $(\theta)$ chosen to be just the generalized spherical 
coordinates of the sphere $S^{n-2}$ and only one parameter of $B_s(\lambda)$.  
These form the angular parameters of $\hat x=\hat x(\theta,\lambda)$ and,
therefore, represent the spherical coordinates $(\theta,\lambda)$ of the
orbit $\Sigma_o$.  Let us observe that, by definition, the generators
$X(\hat x)$ do not depend explicitly on $\lambda$ and, consequently, we can
write
\begin{equation}\label{XXs}
X[\hat x(\theta,\lambda)]=G_o(\theta)X_sG_o(\theta)^{-1}\,.
\end{equation}

In order to fill in the set of the angular coordinates of $\Sigma$, we
introduce other $m$ angular variables
$(\alpha)$ as parameters of ${\bf G}_o$ such that the corresponding
parametrization of ${\bf G}$ should be given by
\begin{equation}\label{parsf}
G(\theta, \lambda, \alpha)=G_o(\theta)B_s(\lambda) G_o(\alpha)\,.          
\end{equation}
This parametrization defines the spherical chart $(\theta, \lambda, \alpha)$
of $\Sigma$ related to the Cartesian one, $(\hat x,y)$, according to 
Eq. (\ref{BGBG}) and the obvious condition $G(\hat x,y)=G(\theta,\lambda,\alpha)$ 
which leads to the rule
\begin{equation}\label{GGG}
G_o(y)=G_o(\theta)G_o(\alpha)\,,
\end{equation} 
giving the coordinates $(y)$ in terms of the angular coordinates 
$(\theta)$ and $(\alpha)$.  We specify that for the matrices
$G_o(y)$ and $G_o(\alpha)$ we can use either the same type or even
completely different types of parametrizations.

The invariant line element (\ref{dsH})  in 
spherical coordinates reads, 
\begin{equation}\label{dsHS}
{ds_h}^2=
-\epsilon_o {\rm Tr}\{P_o\, d\Omega[G_o(\theta)B_s(\lambda)]^2\}\,. 
\end{equation}
This can be evaluated using Eqs. (\ref{OO}) and bearing in mind that 
$X_s$ is fixed so that $d\Omega[B_s(\lambda)]=d\lambda \, X_s$.  
Then, after a few manipulations we find 
\begin{equation}\label{dsHS1}
{ds_h}^2= \epsilon_o\left\{
\begin{array}{lll}
d\lambda^2+\sin^2\lambda \,d\theta^2& {\rm for}
&(\hat x_s\cdot \hat x_o)<1\,,\\
-(d\lambda^2+\sinh^2\lambda \,d\theta^2)& {\rm for}
&(\hat x_s\cdot \hat x_o)>1 \,,
\end{array}\right. 
\end{equation}
where $d\theta^2=-\frac{1}{2}{\rm Tr}[G_o(\theta)^2]$ is  
the standard line element in spherical coordinates of the sphere $S^{n-2}$. 

In the case of the spherical orbits $\Sigma_o$ we can take different
spherical
coordinates for each hemisphere associating the coordinates 
$(\theta,\,\lambda,\,\alpha^{\pm})$  to the Cartesian ones
$(\hat x, y^{\pm})$.
These are related within the transition   
\begin{equation}\label{GHGs}
G_o(\alpha^-)={T}_s(\theta,\lambda)G_o(\alpha^+)\,, 
\end{equation}
which is equivalent to the transition (\ref{GHG}) of the Cartesian charts
if 
\begin{equation}\label{HGHGs}
{T}_s(\theta,\lambda)=
G_o(\theta)^{-1}{T}[\hat x(\theta,\lambda)]G_o(\theta)\,. 
\end{equation}

\section{Isometries and gauge transformations}

As mentioned, one of our major objectives is to find all the manifolds  
(${M}_N, ds)$ of the Kaluza-Klein theories with the isometry group 
${\bf I}(M_N,ds)={\bf G}\otimes {\bf G}_o$ and to identify the Yang-Mills
potentials arising in this context. To this end, we define the isometry
transformations in closed form and  we look for the gauge transformation
of the Yang-Mills potentials produced by these isometries.

\subsection{Isometries and invariant line elements}

In what concerns the isometry transformations, the main point of our 
approach is to postulate how transform the Cartesian coordinates 
$(\hat x, y)$ of $\Sigma$ under the group ${\bf G}\otimes {\bf G}_o$ which has 
to become the isometry group of $\Sigma$ and implicitly of $M_N$.  Since  
the coordinates $(\hat x,y)$ are the parameters of the group ${\bf G}$ we can 
formulate the transformation laws exploiting some special automorphysms  
$(G,G_o):{\bf G}\to {\bf G}$ 
produced by the pairs $(G,G_o)\in {\bf G}\otimes {\bf G}_o$.  First, we 
adopt the passive point of view keeping the manifold fixed and transforming 
among themselves the different charts covering the same domain.  Furthermore, we 
assume that each pair $(G, G_o)$ transforms the Cartesian 
coordinates $(\hat x, y)$ into the new ones $(\hat x', y')$ such that 
the coordinates $(\hat x)$ transform {\em manifestly} covariant as
${\hat x}'=G\hat x$ and
\begin{equation}\label{ISO}
(G, G_o) : G(\hat x, y)\to G({\hat x}',y') =G\, G(\hat x, y)\, G^{-1}_o.    
\end{equation}
Hereby and using Eq. (\ref{Cart}) we obtain the definitive transformation
rules
\begin{eqnarray}
\hat x&\to&{\hat x}'= G\,\hat x \label{trx} \,,\\
G_o(y)&\to& G_o(y')= W(G,\hat x)\,G_o(y)\,G_o^{-1} \,,\label{try}
\end{eqnarray}
where
\begin{equation}\label{W}
W(G,\hat x)=B(G\hat x)^{-1}\, G\, B(\hat x)   
\end{equation}
is a transformation matrix of ${\bf G}_o$ since
$W(G,\hat x)\,\hat x_o =\hat x_o$.
These matrices have a similar structure as the Wigner rotations of
the representation theory of the Poincar\' e group in the momentum space 
\cite{TH}, with the difference that in our case the matrices (\ref{W}) depend
 on coordinates. Consequently, these have similar properties, 
\begin{equation}\label{compW}
W(G'G,\hat x)= W(G',G\hat x)W(G,\hat x)\,,
\end{equation}
and  $W(G,\hat x_o)=1_n$ for any $G\in {\bf G}$ or $W(1_n,\hat x)=1_n$ for
any $x$. 
Hereby we draw the conclusion that  the coordinates $(y)$ transform according
to a representation of the group ${\bf G}$ {\em induced} by the little group 
${\bf G}_o$ \cite{MAK}.
We note that the transformations (\ref{trx}) and (\ref{try}) represent a
{\em combined} transformation in the sense of Ref. \cite{COTA}. These form
a well-defined Lie group with some interesting properties which are presented
in the Appendix B.

The next step is to find the metrics that remain  invariant under these
combined transformations. Since the 1-forms
$d\Omega[G(\hat x, y)]\in {\bf g}$ transform as
\begin{equation}
(G,G_o): d\Omega[G(\hat x, y)]\to d\Omega[G({\hat x}', y')]= 
G_o d\Omega[G(\hat x, y)]G_o^{-1}\,, 
\end{equation}
it is straightforward to show that the invariant metrics are the horizontal 
metric of $\Sigma_o$ given by the line element (\ref{dsH}) and the metric of 
$\Sigma$ defined by Eq. (\ref{sZ}) that in Cartesian coordinates reads
\begin{equation}\label{dsZO}
{ds_Z}^2=-{\rm Tr}\left \{Z\, d\Omega[G(\hat x,y)]^2\right \}\,,
\end{equation}
where $Z$ must be invariant under the transformations of ${\bf G}_o$ 
satisfying
\begin{equation}
G_o Z G_o^{-1}=Z\,,\quad  \forall\, G_o\in {\bf G}_o\,. 
\end{equation}
Therefore, $Z$ is a suitable linear combination of projection operators, 
\begin{equation}\label{ZVH}
Z= \alpha_h P_o+\textstyle\frac{1}{2}\alpha_v I_o\,, 
\end{equation}
involving the arbitrary real constants $\alpha_h$ and $\alpha_v$.  
The invariant metrics (\ref{dsZO}) have specific horizontal and vertical parts 
that can be separated using the formula 
\begin{equation}\label{OHV}
d\Omega[G(\hat x,y)]=G_o(y)^{-1}\left\{-d\Omega[G_o(y)^{-1}]+d\Omega[B(\hat x)]
\right\}G_o(y)
\end{equation}
resulting from Eqs. (\ref{OO}).  We denote by  
\begin{equation}
d\omega(y)=-d\Omega[G_o(y)^{-1}]=[dG_o(y)]G_o(y)^{-1}\, \in {\bf g}_o 
\end{equation}
the vertical 1-form depending on $y$ and split the 1-form $d\Omega[B(\hat x)]$ 
in its horizontal and vertical parts according to Eq. (\ref{split}).  Then we 
calculate the line element (\ref{dsZO}) with the metric operator (\ref{ZVH}) 
using Eqs. (\ref{TrXH}).  The final result is  
\begin{equation}
{ds_Z}^2=\epsilon_o \alpha_h\, {ds_h}^2 + \alpha_v\, {ds_v}^2 \,,
\end{equation} 
where the line element of the vertical metric yields 
\begin{equation}\label{dsVe}
{ds_v}^2=-\textstyle\frac{1}{2} {\rm Tr}\left\{d\omega(y)
+d\Omega[B(\hat x)]_v\right\}^2\,. 
\end{equation}
When $\alpha_h=\alpha_v=1$ the metric operator takes the standard form 
$Z=\frac{1}{2}1_n$.  However, the constants $\alpha_h$ and $\alpha_v$ can be 
replaced at any time by  functions depending only on the invariant 
$|x|$, without to affect the symmetry of the line element. 

These results lead to the conclusion that the manifolds
$(M_N,ds)$ with the isometry group ${\bf I}(M_N,ds)={\bf G}\otimes {\bf G}_o$
must have the Cartesian charts of coordinates $(x,y)$, with
$x^{\mu}=|x|{\hat x}^{\mu}$, where the line elements   
\begin{equation}\label{dsds}
ds^2={ds_o}^2 +f_v(|x|){ds_v}^2\,,\quad {ds_o}^2=f(|x|)(d|x|)^2+
f_h(|x|){ds_h}^2\,,  
\end{equation}
depend on three arbitrary  functions only on $|x|$, denoted by
$f$, $f_h$ and $f_v$.
In order to preserve the original signature of the metric $\eta$
it is recommended to take positive defined functions $f$ and $f_h$.
In this metrization,
the orbits $\Sigma$ and $\Sigma_o$ will get the metrics resulted from the
restriction of the line elements (\ref{dsds}) to fixed $|x|=|x_o|$.

The spherical charts of $(M_N, ds)$, denoted by 
$(|x|,\theta,\lambda,\alpha)$, have a radial coordinate  $|x|$ 
and the angular coordinates $(\theta,\lambda,\alpha)$ of $\Sigma$  
introduced above. The spherical coordinates are related to the Cartesian ones 
according to Eqs. (\ref{BGBG}) and (\ref{GGG}).  Consequently, the horizontal 
metric is given by Eq. (\ref{dsHS1}) while the vertical line element 
(\ref{dsVe}) can be put in the form
\begin{equation}\label{dsVeS}
{ds_v}^2=-\textstyle\frac{1}{2}{\rm Tr}\left\{d\omega(\alpha)+\left[B_s(\lambda)^{-1}
d\Omega[G_o(\theta)]B_s(\lambda)\right]_v\right\}^2\,,
\end{equation}     
where
$d\omega(\alpha)=-d\Omega[G_o(\alpha)^{-1}]=[dG_o(\alpha)]G_o(\alpha)^{-1}$.  In general, the
separation of the vertical part of the second term of  ${ds_v}^2$ is quite 
complicated depending on the concrete choice of $X_s$ and the angular 
variables $(\theta)$.  However, this can be done in each particular 
case separately projecting the entire expression of this term on the  
subalgebra ${\bf g}_o$.   

\subsection{Gauge transformations}

The  manifolds $(M_N, ds)$ with the above isometry properties
have to be considered as the principal fiber bundles of 
the $N$-dimensional Kaluza-Klein theories with central symmetry.
The fibration $(M_N,ds)\to (M_n, ds_o)$ is induced by the fibration
$\Sigma \to \Sigma_o$ so that the fiber bundles $(M_N,ds)$ and $\Sigma$
have the same fiber $F\sim {\bf G}_o$. For this reason the Kaluza-Klein
extra-coordinates
of the charts $(x,y)$ or $(\hat x,y)$ are  the $m$ parameters $(y)$ of 
the little group ${\bf G}_o$.  However, one can introduce other Cartesian
charts,
 $(x,x_K)$ in $(M_N, ds)$ and $(\hat x, x_K)$ in $\Sigma$, with
the new extra-coordinates, $x_K=\phi(y)$, related to the group parameters
through the isomorphism $\phi:{\bf G}_o\to F$. In any case, when these
fiberings are non-trivial we must provide appropriate transition
functions relating the extra-coordinates of the different charts.
 
In the Kaluza-Klein theories the connection on the fiber $F$ is interpreted
as a particular Yang-Mills potential which appears in the vertical
part of the metric defined by the line element (\ref{dsds}). 
We obtain thus physical models where the components of the Yang-Mills
potential, $A_{\mu}\in {\bf g}_o$, are given by the term depending on
$\hat x$ of the vertical metric (\ref{dsVe}).  Consequently, we identify 
\begin{equation}\label{AB}
dA(x)\equiv A_{\mu}(x)dx^{\mu}= {\textstyle\frac{1}{2}}
X_{(\alpha\beta)}A^{(\alpha\beta)}_{\mu}(x)dx^{\mu}=d\Omega[B(\hat x)]_v\,,
\end{equation} 
observing that these potentials are defined up to a real multiplicative
constant  playing the role of the coupling constant in the concrete physical
models. More precisely, when one needs to work explicitly with the
coupling constant $\kappa$ it is enough to rescale $f_v \to \kappa^2 f_v$
and $y\to y/\kappa$ in Eq. (\ref{dsVe}).

A crucial problem is to find the non-Abelian gauge transformations of the 
Yang-Mills field represented by the above particular potentials.  We believe
that in these theories the gauge must have a natural geometrical origin with
a simple physical significance.  Let us observe that the boosts (\ref{Bx})
giving rise to the Yang-Mills potentials (\ref{AB}) are determined up to
an arbitrary transformation of ${\bf G}_o$ depending on coordinates.
For this reason, we say that the section $H : \Sigma_o \to {\bf G}_o$ of the
fiber bundle $\Sigma$ produces the {\em gauge} transformation of the boosts,
\begin{equation}\label{gaugeB}
 B(\hat x)\to B'(\hat x)=B(\hat x)H(\hat x)^{-1}\,,
\end{equation}
generating the gauge transformations \cite{EGH} 
\begin{equation}\label{Aprim} 
 dA(x)\to dA'(x)=H(\hat x)
dA(x)H(\hat x)^{-1}
+d\Omega[H(\hat x)^{-1}]
\end{equation}
of the Yang-Mills potentials defined by Eq. (\ref{AB}).  

The interpretation of the above gauge transformations is simple.
We observe that Eq. (\ref{Aprim}) gives the form of the Yang-Mills potentials 
in the same domain but in a chart corresponding to the new parametrization
$G(\hat x,y')=[B(\hat x) H(\hat x)^{-1}] H(\hat x) G_o(y)\in {\bf G}$.
This is the Cartesian chart $(x,y')$ with the 
new extra-coordinates $y'$ defined by $G_o(y')=H(\hat x)G_o(y)$.  Thus
$H$ plays the role of a transition function relating the
extra-coordinates in the overlapping domain of the charts $(x,y)$ and
$(x,y')$.   

A remarkable property of the Yang-Mills potentials is that their 
transformations under isometries are {combined} with gauge transformations.  
More precisely, whenever one performs the isometry transformations (\ref{trx}) 
and (\ref{try}) then the potentials (\ref{AB}) transform as 
\begin{eqnarray}
&(G,G_o): dA(x)\to dA(Gx)=&W(G,\hat x) dA(x)W(G,\hat x)^{-1} \nonumber\\
&&~~~~~~~~~~+d\Omega[W(G,\hat x)^{-1}]\,,
\label{isoA}
\end{eqnarray}
which means that $A(x)$ behaves like a vector field up to a gauge 
transformation produced by the new transformation matrices (\ref{W})
we defined above.
Therefore, we can say that the potentials (\ref{AB}) have the manifest
central symmetry governed by the group ${\bf G}$.
We observe that our theory of isometries provides correct particular
isometries $(1_n,G_o)$ of the fiber $F$ that do not involve  gauge
transformations.

After we have found the gauge transformations we are able to write down the 
fields strength  
\begin{equation}\label{FAA}
F_{\mu\nu}=\partial_{\mu}A_{\nu}-\partial_{\nu}A_{\mu}+[A_{\mu},A_{\nu}]\,,
\end{equation}
that must satisfy the field equations 
\begin{equation}\label{YMeq}
\nabla^{\mu}F_{\mu\nu}+[A^{\mu},F_{\mu\nu}]=j_{\nu} \,,
\end{equation}   
where $\nabla_{\mu}$ are the first $n$ covariant derivatives of $(M_N,ds)$
and $j_{\nu}\in {\bf g}_o$ is the current of the external sources.
It is known that in the non-Abelian theories the gauge transformations
(\ref{Aprim}) change the form of the fields strength, 
\begin{equation}
F_{\mu\nu}(x)\to F'_{\mu\nu}(x)=H(\hat x)F_{\mu\nu}(x)H(\hat x)^{-1}\,. 
\end{equation}
For this reason the isometry transformations (\ref{isoA}) which involve 
gauge transformations  will transform the 2-forms
\begin{equation}\label{dF}
dF(x)=\textstyle\frac{1}{2}\, F_{\mu\nu}(x)dx^{\mu}\land dx^{\nu}
\end{equation}
according to the rule
\begin{equation}\label{2form}
(G,G_o) : dF(x)\to dF(Gx)=W(G,\hat x)dF(x)W(G,\hat x)^{-1}\,, 
\end{equation}
leading to the conclusion that $F_{\mu\nu}$ transforms like a tensor field
up to a gauge transformation. Hence we can say  that all the models we have
constructed here have a global {\em central} symmetry.

Finally, we specify that the presence of the spinor fields requires to
consider, in addition, the {\em spin} bundle associated to the principal
bundle $(M_N,ds)$. The structure group of the spin bundle is the
{\em gauge group} of the entire theory which must be the universal covering
group $\tilde {\bf G}_o$ of the little group ${\bf G}_o$. In other words, the
gauge group is {\em locally} isomorphic with the little group, both these
groups having the same algebra ${\bf g}_o$. Therefore, the gauge
transformations (\ref{Aprim}) determine the transformations
$\psi(x) \to U(\hat x)\psi(x)$ of the spinor fields whose unitary operators
$U(\hat x)=U[H(\hat x)]\in \tilde {\bf G}_o$ have the same parametrization
as $H(\hat x)$ provided $U(1_n)=1_{sp}$ where  $1_{sp}$ is the identity 
of $\tilde {\bf G}$. 

\subsection{Space-like and time-like models}

Our formalism becomes simpler if we take the string in a suitable fixed 
direction $x^0$ so that $\hat x_o=(1,0,0,...,0)$.
Moreover, we take this direction
of positive signature fixing $\eta_{00}=1$  and $\epsilon_o=1$.  The other 
Cartesian coordinates of $(M_n, ds_o)$ will be denoted by $x^i$,
$i,j,k,. . . =1,2,. . . , n-1$. In what concerns the spherical coordinates,
a convenient choice is  $X_s=X_{(n-1\, 0)}$ and
\begin{equation}\label{GGGG}
G(\theta)=G_s(\theta_1)G_s(\theta_2)....G_s(\theta_{n-2}) \,,
\end{equation}
where $G_s(\theta_j)$ is a rotation of angle $\theta_j$ in the plane
$(x^j,x^{j+1})$, generated by $X_{(j\, j+1)}$. We must specify that the
Euler angles used in the second section is not in accordance to
this parametrization.

By definition the little group ${\bf G}_o$ acts only on the $n-1$ coordinates
$x^i$ leaving the string direction $x^0$ invariant. In order to guarantee the
unitarity of the Yang-Mills theory, the little group must be the
orthogonal group ${\bf G}_o=SO(n-1)$ since then the gauge group will be the
compact group $\tilde{\bf G}_o={\rm Spin}(n-1)$.  This means that the metric
$\eta$ can be either $\eta=1_n$ or $\eta={\rm diag}(1,-1_{n-1})$.  

In the first case when $\eta=1_n$ all the Cartesian coordinates are space-like  
including $x^0$ and we say that the model is {\em space-like} since its 
string is so.  Obviously, all the space-like models are {\em static} such that 
physical models can be constructed only adding the time in a trivial manner.
All the fields strength of these models are of {\em magnetic} type since
$(M_n,ds_o)$ has no time-like coordinates. 

However, when  $\eta={\rm diag}(1,-1_{n-1})$ we have one time coordinate $x^0$
and $n-1$ space coordinates $x^i$ but the string is along the time direction 
$x^0$.  The model with a string of this type will be called {\em time-like} 
model.  In these models we have to meet fields of both electric and magnetic
types. Interesting candidates for these models are the geometries having
the same $SO(3,1)$ isometries as the Minkowski spacetime.   
      
For the both types of models the matrices $X_{(ij)}$ are the {\em vertical}
basis-generators of the algebra ${\bf g}_o=so(n-1)$ while the generators
$X_{(i0)}$ span the coset space ${\bf g}/{\bf g}_o$.  In this context the
definition (\ref{AB}) yields
\begin{equation}\label{Aij}
A_k=-\frac{X_{(kj)} x^j}{|x|(|x|+x^0)}\,, \quad
A_0=0\,. 
\end{equation}   
These potentials are defined up to  gauge transformations that may
dramatically change this very simple form.  Obviously, the difference between
the space-like and time-like models is encapsulated in the expression of
$|x|$ as well as in the physical meaning of the coordinate $x^0$. 

Each space-like or time-like model of a given dimension is determined by
the concrete form of its invariant functions $f$, $f_h$ and $f_v$ and by the
fibration that can be chosen independently on these functions.  This means that
there are many models with the same symmetry and Yang-Mills potentials
(up to a gauge) but with {\em different} geometries.  For this reason it is
useful to divide the set of all of these models in {\em classes} of symmetry. 
We say that all the space-like models with manifolds $(M_N, ds)$ of dimension
$N=\frac{1}{2}n(n-1)+1$ and
isometry group ${\bf I}(M_N,ds)=SO(n)\otimes SO(n-1)$ form the class
of symmetry ${\cal S}(n)$. The time-like models of the same dimension
but with the isometry group ${\bf I}(M_N,ds)=SO(1,n-1)\otimes SO(n-1)$ 
constitute the class of symmetry ${\cal S}(1,n-1)$.  

The time-like models are completely new and seem to be rather special
since these are no static and have extended singularities on the light-cone.
For this reason we believe that these will rise new delicate problems that
may be carefully analyzed elsewhere. In what follows we restrict  ourselves
to the space-like models of the classes ${\cal S}(n)$ which will be
interpreted as generalized monopoles. 

\section{Monopole models}

In the space-like models of the class ${\cal S}(n)$ the base manifolds
$(M_n,ds_o)$ have only space-like Cartesian coordinates. Then the metric
is $\eta=1_n$, the group ${\bf G}=SO(n)$ is orthogonal and the forms
$(x\cdot x)=|x|^2=(x^0)^2 + \eta_{ij}x^ix^j$ are positively defined.
This justifies the usual  notation  $r=|x|$ for
the {\em radial} coordinate of the spherical chart of $(M_N,ds)$ denoted
from now by $(r,\,\theta,\,\lambda,\,\alpha)$. Here the line element
is of the general form (\ref{dsds}) where $d{s_h}^2$ is given by the  first
of Eqs. (\ref{dsHS1}) while the vertical part has to be calculated according  
to Eq. (\ref{dsVeS}). In the Cartesian coordinates $(x,y)$ of the
same manifold, the line element resulted from  Eqs. (\ref{dsHo}), 
(\ref{dsVe}) and (\ref{AB}) is
\begin{equation}\label{dsCC} 
ds^2=f(r)(\hat x\cdot dx)^2+f_h(r)(d\hat x\cdot d\hat x)
- \textstyle\frac{1}{2}f_v(r) {\rm Tr}\left[d\omega(y)
+ A_{\mu}(x)dx^{\mu} \right]^2\,. 
\end{equation}
Let us observe that in our notation the Euclidean flat metric can be written
as ${ds_E}^2=(dx\cdot dx)= (\hat x\cdot dx)^2 +r^2\, (d\hat x\cdot d\hat x)$.
In general, the geometry of $(M_N,ds)$ is determined only by the choice of
the invariant functions since the form of the potentials is given by
Eq. (\ref{AB}). The  most interesting case is when one can match these
functions so that the metric of $(M_N,ds)$ be an exact solution of the vacuum
Einstein equations.

\subsection{The monopole potentials}

The orbits $\Sigma_o \sim S^{n-1}$ of the  models $(M_N,ds)\in {\cal S}(n)$
are spheres of fixed radius $r$.  Here we have the pair 
of charts $(x,y^{\pm})$ of the hemispheres $\Sigma_o^{\pm}$ 
where the Yang-Mills potentials may take different forms  $A^{\pm}$. 
Obviously, on the domain $\Sigma_o^+\bigcap \Sigma_o^-$ these must differ
among themselves only within a gauge.  Therefore, the potentials
$A^{\pm}$  must be related to each other through  
a suitable gauge transformation (\ref{Aprim}) where if $A=A^+$ then $A'=A^-$.
 In order to construct an efficient mechanism of this type, it is indicated
to consider the gauge transformation (\ref{isoA}) associated to isometries 
rather than looking for arbitrary transition functions. 

The group $O(n)$ has many proper or improper transformations able to change
the sign of the axis $x_o$.  All
of them  can be used for defining the gauge we need for our fiberings but
here we consider only the proper transformations. Let us denote by
$Q\in SO(n)$ a proper matrix that performs $Q x_o=-x_o$ and take the
potentials of the upper hemispheres
\begin{equation}\label{Aplus}
A_k^{+}(x)=-\frac{X_{(ki)}x^i}{r(r+x^0)}\,, \quad A_0=0\,.   
\end{equation}
Then the potentials of the lower hemispheres must be of the form  
\begin{equation}\label{Aminus}
A_\mu^{-}(x)=Q_{\cdot\, \mu}^{k\,\cdot}A_k^{+}(Qx) 
=-\frac{X_{(ki)}Q_{\cdot\,\mu}^{k\,\cdot}Q^{i\,\cdot}_{\cdot\, \nu}x^{\nu}}
{r(r-x^0)}\,,   
\end{equation}
since these differ from $A_k^+$ only through the gauge transformation 
(\ref{isoA}) that now reads
\begin{equation}\label{AQH}        
dA^{-}(x)\equiv dA^{+}(Q x)={T[Q]}(\hat x)dA^{+}(x){T[Q]}(\hat x)^{-1}+
d\Omega[{T[Q]}(\hat x)^{-1}]\,,
\end{equation}
where ${T[Q]}(\hat x)=W(Q,\hat x)$ can be calculated from
Eqs.  (\ref{W}) and (\ref{BBBQ}) obtaining
\begin{equation}\label{cuiu}
{T[Q]}(\hat x)= Q[1_n+2X(\hat x)^2]\,.
\end{equation}
The mapping $T[Q]: \Sigma_o \to SO(n-1)$  is the transition function of
the fibration that provides a suitable topological structure of the principal
bundle $(M_N,ds)$,  reducing the effects of the string singularity up to a
point-like one. We note that the matrices $Q$ are defined up to left or/and
right multiplications with arbitrary $SO(n-1)$ matrices that leave $x_o$
invariant.
Each matrix $Q$ determines its specific fibration such that we have many
possibilities of choice. However, these must be equivalent since the theory
has a central symmetry.

In the spherical coordinates defined by Eqs. (\ref{parsf})
and (\ref{GGGG}), the transition  (\ref{GHGs})
is produced by an $SO(n-1)$ matrix which, according to
Eqs. (\ref{XXs}) and (\ref{HGHGs}),
depends only on the angular coordinates $(\theta)$ as
\begin{equation}\label{Htheta}
{T[Q]}_s(\theta)=G_o(\theta)^{-1}QG_o(\theta)B_s(\pi)\,. 
\end{equation}
This formula becomes simpler if we take $Q=B_s(\pi)=1_n+2X_s$ since then 
$B_s(\pi)$ commutes with the rotations $G_s(\theta_j)$ of Eq. (\ref{GGGG})
for all $j=1,2,...,n-3$ while for $j=n-2$ we have $B_s(\pi)G_s(\theta_{n-2})
B_s(\pi)=G_s(-\theta_{n-2})$. Therefore, we find that the transition
functions of the spherical charts have the values
\begin{equation}\label{Htheta1}
{T[Q]}_s(\theta)=G_s(-2\theta_{n-2})\,, 
\end{equation}
for $0\le \theta_{n-2}<2\pi$.

This result shows us that for all our space-like models the transition
functions are simple rotations in the plane $\{x^{n-1},x^{n-2}\}$ orthogonal
to the string direction $x^0$. This means that the {\em non-trivial} part of
the principal bundle $\Sigma$ reduces to the bundle over the two-dimensional
sphere embedded in the Euclidean subspace $\{x^0, x^{n-1}, x^{n-2}\}$
where the transition function (\ref{Htheta1}) provides $SO(2)$ rotations
defined on the equatorial circle.
Because of the central symmetry, this result can be reproduced for any other
two-dimensional sphere $S^2_{ij}$ which surrounds the singularity at $x=0$,
being embedded in the Euclidean subspace $\{x^0,x^i,x^j\}$, $i\not = j$.
For this purpose it suffices to take an appropriate fibration whose
transition function should be a rotation in the plane $\{x^i,x^j\}$.

In other respects, it is remarkable that the transition functions are defined
on the angular domains $[0, 4\pi)$ that cover twice the equatorial circles.
This assures the correct gauge transformations of the spinor fields whose
gauge groups are simply connected. More precisely, the gauge transformations
of the spinors corresponding to the rotations (\ref{Htheta1}) are produced by
the unitary transformations $U(-2\theta_{n-2})\in {\rm Spin}(n-1)$ which
satisfy $U(-4\pi)=1_{sp}$ when $\theta_{n-2}=2\pi$. Otherwise, if this
domain would be only $[0,-\pi)$ then the transformation $U(-2\pi)=-1_{sp}$
might change the sign of spinors, the theory becoming thus pointless \cite{WY}.

Hereby we conclude that the remaining point-like singularity behaves as a
{\em monopole} since it is topologically {\em stable} \cite{BPS2} and,
therefore, the topology of the principal bundle is of the homotopy type
$\pi_1[SO(n-1)]= {\Bbb Z}_2$ if $n>3$. Thus the string singularities of the
Yang-Mills potentials of all our non-Abelian models can be reduced up to
monopoles in a similar manner as in the Abelian case of the Dirac magnetic
monopole in the Taub-NUT background which is of the homotopy type
$\pi_1[SO(2)]=\pi_1[U(1)]={\Bbb Z}$ since $n=3$.

\subsection{The fields strength}

For understanding the behavior of the field represented by the above
potentials, let us calculate the fields strength (\ref{FAA}). In the case of
the non-Abelian gauge theories the gauge transformations change the form of
the fields strength without to affect the physical meaning of the theory. For
this
reason  the potentials $A^{\pm}_{\mu}$ give rise to the different fields
strength $F^{\pm}_{\mu\nu}$ related to each other within the gauge
transformation (\ref{AQH}) which yields
\begin{equation}
 dF^{-}(x)\equiv dF^{+}(Qx)={T[Q]}(\hat x)dF^{+}(x){T[Q]}(\hat x)^{-1}\,. 
\end{equation}
Consequently, it will suffice to calculate the components of
the field $F_{\mu\nu}\equiv F^{+}_{\mu\nu}$ in terms of
$A_{\mu}\equiv A_{\mu}^+$ as
\begin{eqnarray}
F_{0i}(x)&=&X_{(ik)}\frac{x^k}{r^3}\,,\label{FF1}\\
F_{ij}(x)&=&\frac{1}{r^2}\left[X_{(ij)}+A_i(x)x^j-A_j(x)x^i\right]\,.
\label{FF2}
\end{eqnarray}
One can verify that these fields strength are {\em exact}
solutions of the sourceless Yang-Mills equations (\ref{YMeq}). Starting
with the obvious identities
$x^{\mu}A_{\mu}=0$ and $x^{\mu}F_{\mu\nu}=0$ one finds $j_0=0$ 
which is compatible with the global $SO(n)$ symmetry only
when $j_{\nu}=0$. Thus the requirements (I)-(III) are fulfilled remaining
to study only the invariants.

The main differential forms related to the fields strength are the 2-form
(\ref{dF}) and the {\em dual} $(n-2)$-form
\begin{equation}\label{dFdual}
dF^{*}(x)
=\frac{1}{2(n-2)!}\,F_{\alpha\beta}(x)\,
\varepsilon^{\alpha\beta}_{\cdot\,\cdot\,\,\sigma_1 ... \sigma_{n-2}}
 dx^{\sigma_1}\land dx^{\sigma_2}\land ... dx^{\sigma_{n-2}}\,,
\end{equation}
where $\varepsilon_{\alpha_1\alpha_2...\alpha_n}$ is the
total skew-symmetric $SO(n)$ tensor. With the help of $dF$ and $dF^*$ we can
define the following candidates of invariants
\begin{equation}\label{flux}
\Phi(S^2) =\int_{S^2} dF\,,\qquad
\Phi^*(S^{n-2}) =\int_{S^{n-2}} dF^*\,.
\end{equation}
These integrals are elements of the $so(n-1)$ algebra representing
fluxes through two-dimensional and, respectively, $(n-2)$-dimensional spheres
of arbitrary radius $r_0$. 
We observe that the value of the first integral does not depend on $r_0$
so that the integration can be done over the unit sphere. 

The orbit $\Sigma_o= S^{n-1}$ of unit radius
embeds a number of $\frac{1}{2}(n-1)(n-2)=m$ different unit spheres
$S^2_{ij} \subset \Sigma_o$ surrounding the monopole. Therefore, we shall
obtain $m$ different corresponding
values of $\Phi$, each one depending on the $m$ basis-generators
$X_{(ij)}$ of the $so(n-1)$ algebra. In this manner we generate a real
$m\times m$ matrix of flux components that can be put in diagonal form.
Indeed, if we consider the complete system of $m$ spheres
$S_{ij}^2$, $i,j=1,2,..., n-1$  $(i\not = j)$, then a simple calculation
leads to the interesting result
\begin{equation}\label{PhiX}
\Phi(S_{ij}^2)=4\pi X_{(ij)}
\end{equation}
that holds for any space-like model. Since the generalized monopoles are
topologically stable these fluxes represent the principal invariants of
our space-like models.

In what concerns the second integral of (\ref{flux})
we observe that the number of spheres $S^{n-2}\subset S^{n-1}$ is $n-1$. 
In general, if $n\not=4$ this number differs from $m$ which means that
the number of integrals $\Phi^*\in so(n-1)$ differs from that of the
basis-generators of the $so(n-1)$ algebra. This fact is not in agreement
with the $SO(n-1)$ gauge symmetry, indicating  that the fluxes
$\Phi^*(S^{n-2})$ may vanish. In the particular case of $n=4$ the
concrete calculation leads to the same result as we shall see in the
Example 2 of the next section. 

For constructing second order invariants we can
not use the 4-form $dF\land dF$ since this is traceless. For this reason,
we must consider the integrals  
\begin{equation}\label{Theta}
\Theta(V) =-\frac{1}{2}{\rm Tr}\int_{V}
dF(x)\land dF^*(x)\,,
\end{equation}
over arbitrary volumes $V\subset M_n$. The singularity at $x=0$ of the
fields strength makes these integrals divergent for the models with $n<5$ but
if $n\ge 5$ and $M_n$ is a compact manifold then the quantity $\Theta(M_n)$
is a finite real number that may be interpreted as a second order 
invariant since, in general, this does not vanish.
 
The general conclusion is that the potentials (\ref{Aplus}) and (\ref{Aminus})
give rise to a field strength with $SO(n)$ central symmetry  which is a
particular  solutions of the sourceless Yang-Mills equations in the
$n$-dimensional physical space of a Kaluza-Klein theory with the gauge
group ${\rm Spin}(n-1)$ and $SO(n)\otimes SO(n-1)$
isometries. It is remarkable that this solution is topologically stable
allowing as principal invariants the fluxes of the fields strength
through the two-spheres surrounding the singularity at $x=0$.
Since the values of these fluxes are $4\pi$, in units of coupling
constant, we can say that the space-like models presented here
have the properties (I)-(IV) which, in our opinion, define a plausible
version of generalized monopoles.
 
\subsection{Examples}

Our method of finding monopole solutions gives rise to a large collection
of models divided in classes of symmetry ${\cal S}(n)$. Each of these classes
is formed by all the models which have the same symmetry and implicitly the
same gauge group. We believe that
some of them could be attractive as can be seen
from the table below which lists the first five classes of symmetry pointing
out the dimensions, gauge groups \cite{GG}  and fibrations.

\begin{center}
\begin{tabular}{|c|c|c|c|}
\hline
symmetry&dimension&gauge group&fibration\\
class ${\cal S}(n)$&$(M_N,ds)$&Spin$(n-1)$&$\Sigma\to \Sigma_o$\\
\hline
$n=3$&$N=4$&$U(1)$&$S^3\to S^2$\\
\hline
$n=4$&$N=7$&$SU(2)$&$S^6\to S^3$\\
\hline
$n=5$&$N=11$&$SU(2)\otimes SU(2)$&$S^{10}\to S^4$\\
\hline
$n=6$&$N=16$&$Sp(2)$&$S^{15}\to S^5$\\
\hline
$n=7$&$N=22$&$SU(4)$&$S^{21}\to S^6$\\
\hline
\end{tabular}
\end{center}

In general, the models of a given class of symmetry have the same monopole
fields but different geometries  of the principal bundle $(M_N,ds)$
determined by the
invariant functions $f$, $f_h$ and $f_v$. This offers one the opportunity to
find new models that could be of some physical interest. Of course, first
we look for the monopole models whose manifolds
should be  new solutions of the Einstein equations without matter terms.
In what follows we restrict ourselves to discuss a well-known geometry and to
present examples of new $SU(2)$ models with Einstein metrics.

\subsubsection*{Example 1: The Euclidean Taub-NUT space} 

A simple but famous example is the $U(1)$  Dirac magnetic monopole in the 
Euclidean Taub-NUT space.  This is a special member of the class of symmetry
${\cal S}(3)$ we have presented in the second section as argument for our
attempt. 

This has the virtue to be Ricci flat, its metric being an exact solution
of the Einstein equations in vacuum. In the Cartesian charts
$(\vec{x},y^{\pm})$ the line elements yield
\begin{equation}\label{dsTNUT}
{ds_{\pm}}^2=\frac{1+r}{r}\,d\vec{x}\cdot d\vec{x}+
\frac{r}{1+r}(dy^{\pm}+A_i^{\pm}dx^i)^2   \,,
\end{equation}
where $A^{\pm}$ are the potentials of the Dirac magnetic monopole. Since the
string is along the third axis we have $x^0=x^3$ and only one vertical
generator, $X_{(12)}$. The potentials are defined by Eqs. (\ref{Aplus})
and (\ref{Aminus}) where we chose
$Q=B_s(\pi)=R_2(\pi)={\rm diag}(-1,1,-1)$  finding the components 
\begin{equation}
A_1^{\pm}=\mp \frac{x^2}{r(r\pm x^3)}\,, \quad
A_2^{\pm}=\pm \frac{x^1}{r(r\pm x^3)}\,, \quad
A_3^{\pm}=0\,. 
\end{equation}
These give rise to the magnetic field with central symmetry
\begin{equation}
\vec{B}={\rm rot}\vec{A}^{\pm}=\frac{\vec{x}}{r^3}\,. 
\end{equation}

In spherical coordinates, the set $(\theta)$ reduces to the angle $\varphi$,
the angle $\theta$ plays the role of $\lambda$  and the only spherical
extra-coordinate is $\alpha$. Therefore, we replace $G_o(\theta)$ with
$R_3(\varphi)$ and $B_s(\lambda)$ with $R_2(\theta)$ taking
$G_o(\alpha)=R_3(\alpha)$. Calculating the line elements according to our
general method we obtain the well-known result
\begin{equation}
{ds_{\pm}}^2=\frac{1+r}{r}\,(dr^2+r^2d\theta^2 + r^2 \sin^2\theta d\varphi^2)
+\frac{r}{1+r}(d\alpha^{\pm}+ \cos\theta d\varphi)^2 \,,
\end{equation}
that may be derived directly from Eq. (\ref{dsTNUT}).  
Here the spherical coordinates $\alpha^{\pm}$ are related to each other
according to Eq. (\ref{GHGs}) whose  transition function 
$(\ref{Htheta1})$ is now ${T_s}(\varphi)=R_3(-2\varphi)$
so that $\alpha^-=\alpha^+ -2\varphi$,
recovering thus a version of the Hopf fibration $S^3\to S^2$.

\subsubsection*{Example 2: $SU(2)$  models}

Let us consider the case of $n=4$ and $N=7$ when the Kaluza-Klein
manifolds $(M_7,ds)\in {\cal S}(4)$ are principal fiber
bundles whose base manifolds, $(M_4, ds_o)$, have the associated flat metric
$\eta=1_4$. In these models the little group is $SO(3)$ and the gauge group
is $SU(2)$.

In general, a chart $(u)$ of $(M_7,ds)$ has the
coordinates $u^A$, $A,B,...=0,1,...,6$. The Cartesian coordinates $(x,y)$ are
formed by the physical ones,
$(x)=(x^0, \vec{x})$, and the arbitrary $SU(2)$ parameters, $(y)$. The
coordinates $\vec{x}$ span the subspace $M_3\subset M_4$ orthogonal to the
direction $x^0$. Since $M_3$ has the $SO(3)$ symmetry we can use here the
vector notation taking into account that now $i,j,...=1,2,3$. The
corresponding spherical coordinates
$(r,\,\theta,\,\varphi,\,\lambda)$
of the space $(M_4,ds_o)$ are defined by the transformation
$x=G_o(\theta)B_s(\lambda)x_o$ with $G_o(\theta)\equiv R(\varphi,\theta, 0)$
while $B_s(\lambda)$ is a rotation of angle $\lambda$ in the plane
$\{x^0,x^3\}$. Consequently, we have
\begin{eqnarray}
x^0&=&r\cos\lambda\,,\nonumber\\
x^1&=&r\sin\lambda\sin\theta\cos\varphi\,, \label{cart} \\
x^2&=&r\sin\lambda\sin\theta\sin\varphi\,,\nonumber\\
x^3&=&r\sin\lambda\cos\theta\,. \nonumber
\end{eqnarray}
In addition, we introduce three spherical Kaluza-Klein extra-coordinates 
$(\alpha)$ denoted from now by $\alpha^1=\alpha,\,\alpha^2=\beta$ and
$\alpha^3=\gamma$ which represent the Euler parameters of the rotations 
$G_o(\alpha)\equiv R(\alpha,\beta, \gamma)\in SO(3)$. 

Following the general method, we start with the spherical charts of $(M_7,ds)$ 
having the coordinates $(u_{\pm})\equiv (r,\theta,\phi, \lambda,
\alpha^{\pm},\beta^{\pm},\gamma^{\pm})$ and the line elements
\begin{eqnarray}
{ds_{\pm}}^2&=&g_{AB}(u_{\pm})du_{\pm}^Adu^B_{\pm} \nonumber\\
&=&f(r)dr^2 
+f_h(r)[d\lambda^2+\sin^2\lambda(d\theta^2+\sin^2\theta d\varphi^2)]\nonumber\\
&&+ f_v(r)\left\{(d\alpha^{\pm}+\cos\beta^{\pm} d\gamma^{\pm} +
\cos\theta d\varphi)^2\right. \nonumber\\
&&+(\sin\alpha^{\pm}\sin\beta^{\pm} d\gamma^{\pm} +\cos\alpha^{\pm}
d\beta^{\pm}+ \cos\lambda d\theta )^2\nonumber\\
&&\left. +(\cos\alpha^{\pm}\sin\beta^{\pm} d\gamma^{\pm}
-\sin\alpha^{\pm} d\beta^{\pm}-
\cos\lambda \sin\theta d\varphi)^2
\right\}\,.
\end{eqnarray}
Furthermore, we consider the non-trivial fibration defined by the transition
function (\ref{Htheta1}) that in this case yields
$T[Q]_s(\theta)=R_2(-2\theta)$ since $Q=B_s(\pi)={\rm diag}(-1,1,1,-1)$.  
The resulted fiber bundle $(M_7,ds)$ is non-trivial only on the sphere
$S^2_{13}$. The transition transformation
\begin{equation}
R(\alpha^-,\beta^-,\gamma^-)=
R_2(-2\theta)R(\alpha^+,\beta^+,\gamma^+)
\end{equation}
can be solved in terms of the Euler variables but the result is quite
complicated.

The geometry of the principal bundles $(M_7,ds)$ is determined by the
functions $f$, $f_h$ and $f_v$. Looking for models with 
metrics satisfying the vacuum Einstein
equations, we find two solutions of the positive curvature
$R_{AB}=6k^2g_{AB}$,
\begin{eqnarray}
&{\rm sol.~ 1 :}& f(r)=3\,,\quad f_h(r)=\frac{1}{k^2}\sin^2 kr\,,
\quad f_v(r)=\frac{1}{3k^2}\sin^2 kr\,,\nonumber\\
&{\rm sol.~ 2 :}& f(r)=5\,,\quad f_h(r)=\frac{1}{k^2}\sin^2 kr\,,
\quad f_v(r)=\frac{1}{k^2}\sin^2 kr\,,\nonumber
\end{eqnarray}
and two solutions of the negative curvature
$R_{AB}=-6k^2g_{AB}$,
\begin{eqnarray}
&{\rm sol.~ 1 :}& f(r)=3\sec^2 kr\,,\quad f_h(r)=\frac{1}{k^2}\tan^2 kr\,,
\quad f_v(r)=\frac{1}{3k^2}\tan^2 kr\,,\nonumber\\
&{\rm sol.~ 2 :}& f(r)=5\sec^2 kr\,,\quad f_h(r)=\frac{1}{k^2}\tan^2 kr\,,
\quad f_v(r)=\frac{1}{k^2}\tan^2 kr\,,\nonumber
\end{eqnarray}
where $k$ is a real constant. In both the above cases the limit $k\to 0$
leads to the Ricci flat metrics with $R_{AB}=0$,
\begin{eqnarray}
&{\rm sol.~ 1 :}& f(r)=3\,,\quad f_h(r)=r^2\,,
\quad f_v(r)=\textstyle\frac{1}{3}r^2\,,\nonumber\\
&{\rm sol.~ 2 :}& f(r)=5\,,\quad f_h(r)=r^2\,,
\quad f_v(r)=r^2\,.\nonumber
\end{eqnarray}
All these results have been obtained using Maple.

In the Cartesian charts $(x,y^{\pm})$ these solutions give the line elements
of the form (\ref{dsCC}) with the Yang-Mills potentials (\ref{Aplus})
and, respectively, (\ref{Aminus}).
The above fibering joints the descriptions of the gauge field in both these
charts such that the physical effects depend only on the fields strength 
(\ref{FF1}) and (\ref{FF2}). These can be written in the vector notation
if we use the $SO(3)$ generators $J_i=\frac{1}{2}\varepsilon_{ijk}X_{(jk)}$
and we put
\begin{equation}
E_i=F_{0i}\,,\qquad B_i=\textstyle\frac{1}{2}\varepsilon_{ijk}F_{jk}\,.
\end{equation}
Then we obtain
\begin{equation}\label{EB}
\vec{E}=\frac{\vec{x}\times \vec{J}}{r^3}\,,\qquad
\vec{B}=\frac{x^0\,\vec{J}}{r^3}+
\frac{\vec{x}\,(\vec{x}\cdot \vec{J})}{r^3(r+x^0)}\,,
\end{equation}
where $r^2=\vec{x}^2+(x^0)^2$.

In this 4-dimensional model there exists the {\em dual} field strength 
$F_{\mu\nu}^* =\frac{1}{2}\varepsilon_{\mu\nu\alpha\beta}F^{\alpha\beta}$
giving the 2-form $dF^*$ defined by Eq. (\ref{dFdual}). The 2-forms $dF$ and
$dF^*$ help us to calculate the integrals (\ref{flux}) recovering the result
(\ref{PhiX}) and $\Phi^*=0$ as was expected. In this model the integral
(\ref{Theta}) is divergent whereas other second order invariants can not be
constructed since ${\rm Tr}(dF\land dF)={\rm Tr}(dF^*\land dF^*)=0$.

Even though it is premature to draw definitive conclusions, we observe that
the values of these invariants as well as the  fields (\ref{EB})
indicate that our $SU(2)$  models are new. It seems that these are closer to
the BPS monopoles \cite{BPS} rather than other models of $SU(2)$
monopoles \cite{YA,TP,MIN}.

\section{Concluding remarks}

Here we have shown that there are non-Abelian Kaluza-Klein geometries
in $N$-dimensional principal fiber bundles with $SO(n)\otimes SO(n-1)$
isometries,
having $n$-dimensional bases with manifest $SO(n)$ central symmetries and
fibers $F\sim SO(n-1)$ representing the little groups the string directions.
The Yang-Mills potentials (\ref{Aij}) appear as being produced by strings but
the string singularities can be reduced up to a point-like ones since the
non-trivial principal bundles have suitable topological properties. Thus we
obtain theories where the surviving singularities are topologically stable
and, therefore, can be interpreted as  generalized monopoles as long as the
conditions (I)-(IV) are fulfilled. As a matter of fact, our approach is an
inverse method which recovers the Yang-Mills fields from a given Kaluza-Klein
geometry, avoiding to solve directly the Yang-Mills equations.

A remarkable result concerns the time-like models that are new
to our knowledge. The Yang-Mills fields of these models are produced by
time-like strings in physical backgrounds having one time-like coordinate
and arbitrary space-like ones.  The models of this type are no static and
their Yang-Mills potentials are singular on the future light cone when the
string is in the past. In this situation it is clear that the effects of this
extended singularity can not be removed neither within topology nor using
other methods. Thus we found the new classes of symmetry ${\cal S}(1,n-1)$
whose models seem to be rather special, with new and unusual properties. We
hope that further investigations should clarify if these models could have
a physical meaning.

From the technical point of view the central points of our approach are the
parametrization (\ref{Cart}) written with the help of the boosts (\ref{Bx})
and the integral form of the isometry
transformations (\ref{ISO}) we postulated here. These lead to the induced
representations (\ref{try}) giving rise to the associated gauge
transformations we needed for constructing  models with central symmetry and
the specific transition functions defining our fiberings. As observed
in the section 4, these induced representations arise from an orbital
analysis similar to the well-known one of the representation theory of the
Poincar\' e group in the momentum space. The difference is that our induced
representations are connected to the Kaluza-Klein geometry since
their transformations depend on coordinates producing  gauge transformations.
This indicates that the study of these representations could emphasize new
interesting mathematical properties.

Another mathematical challenge is how could be extended the above presented
method of induced representations to theories with other types of
singularities the reduction of which should require bundles of higher
homotopy types, $\pi_k$ with $k>1$ \cite{YY}. A solution could be to
replace the strings of our monopole models by branes.

\subsection*{Acknowledgments}

I should like to thank Mihai Visinescu for helpful discussions and
suggesting me to look for new monopole models with Einstein metrics. 
This work is partially supported by MEC-AEROSPATIAL Program, Romania.

\appendix

\section{The generators $X(\hat x)$}

The real basis-generators $X_{(\mu\nu)}$ we use here have the matrix elements 
\begin{equation}
\left<X_{(\alpha\beta)}\right>^{\mu\,\cdot}_{\cdot\, \nu}=
\delta^{\mu}_{\alpha}\eta_{\beta\nu}-\delta^{\mu}_{\beta}\eta_{\alpha\nu}\,,
\end{equation}
and the trace properties
\begin{equation}
{\rm Tr}\left(X_{(\mu\nu)}\right)=0\,,\quad
{\rm Tr}\left(X_{(\mu\nu)}X_{(\alpha\beta)}\right)= 2\left(
\eta_{\mu\beta}\eta_{\nu\alpha}-
\eta_{\nu\beta}\eta_{\mu\alpha}\right)\,,
\end{equation}
allowing one to define the Killing forms.
They satisfy the commutation relations
\begin{equation}
\left[ X_{(\mu\nu)},X_{(\sigma\tau)}\right]=
\eta_{\mu\tau} X_{\nu\sigma}
-\eta_{\mu\sigma} X_{\nu\tau}
+\eta_{\nu\sigma} X_{\mu\tau}
-\eta_{\nu\tau} X_{\mu\sigma}\,,
\end{equation}
and transform as
\begin{equation}
GX_{(\mu\nu)}G^{-1}=G^{\cdot\,\alpha}_{\mu\,\cdot}
G^{\cdot\,\beta}_{\nu\,\cdot}X_{(\alpha\beta)}\,. 
\end{equation}
The boost generators defined by Eq. (\ref{XH}) have the obvious properties
\begin{equation}
X(\hat x)^T=sX(\hat x)\,, \quad X(\hat x)^3=s X(\hat x)\,, \quad 
{\rm Tr}[X(\hat x)^2]=2s\,,  
\end{equation}
where $s={\rm sign}[(\hat x\cdot \hat x_o)^2-1]$. 
In the unitary case when $(\hat x\cdot \hat x_o)=\cos\lambda$, the
transformations $Q\in SO(n)$ that satisfy $Q x_o=-x_o$ change
$\lambda \to \pi-\lambda$ and give 
$X(Q\hat x)=-QX(\hat x)Q^{-1}$ such that
\begin{equation}\label{BBBQ}
B(Q\hat x)=Q[1_n+X(\hat x)^2(1+\cos\lambda)-X(\hat x)\sin\lambda]Q^{-1}\,. 
\end{equation}
Hereby we find the important result (\ref{cuiu}).

\section{Induced representations}

Let us start with the principal bundle $\Sigma$ with the fiber
$F\sim {\bf G}_o$ and the base manifold $\Sigma_o$ where we choose
the Cartesian chart $(\hat x)$. We denote by ${\rm Sec}(\Sigma)$
the set of the (local) sections $V: \Sigma_o \to {\bf G}_o$ that form a
group with respect to the multiplication $\,\times\,$ defined as
$(V'\times V)(\hat x)=V'(\hat x)V(\hat x)$ for all $V,V'\in {\rm Sec}(\Sigma)$. The
identity section obeys ${\rm Id}(\hat x)=1_n$ and the inverse section of $V$
is $V^{-1}$ so that  $V^{-1}(\hat x)=[V(\hat x)]^{-1}$. For any automorphism
$\phi : \Sigma_o \to \Sigma_o$ we can define the sections
$V\circ \phi$ with the obvious properties ${\rm Id}\circ \phi ={\rm Id}$,
$V\circ {\rm id}=V$, when $\phi={\rm id}$ is the mapping identity, and
$(V'\times V)\circ \phi =(V'\circ \phi)\times (V\circ \phi)$.

We denote by ${\rm Aut}(\Sigma_o)$ the group of automorphisms of $\Sigma_o$
and we say that the pair $(V,\phi)$, with $V\in {\rm Sec}(\Sigma)$ and
$\phi \in {\rm Aut}(\Sigma_o)$, represents a {\em combined} transformation. 
These pairs constitute a group ${\cal G}$ with respect to the
multiplication $\,*\,$ defined as follows:
\begin{equation}
(V',\phi')*(V,\phi)=((V'\circ\phi)\times V, \phi'\circ\phi)\,.
\end{equation}
The identity of this group is $($Id,id$)$ while the inverse of a pair
$(V,\phi)$ reads
\begin{equation}
(V,\phi)^{-1}=(V^{-1}\circ\phi^{-1}, \phi^{-1})\,.
\end{equation}
One can verify that ${\cal G}={\rm Sec}(\Sigma) {\,\circledS\,} {\rm Aut}
(\Sigma_o)$
is a semidirect product where ${\rm Sec}(\Sigma)$ is the invariant subgroup
\cite{COTA}.

The isometries of $\Sigma_o$ define the set of mappings $\phi_G$ as
$\hat x\to{\hat x}'=G\hat x\equiv \phi_G(\hat x)$ for all $G\in {\bf G}$.
These form a group since $\phi_G'\circ\phi_G=\phi_{G'G}$,
$(\phi_G)^{-1}=\phi_{G^{-1}}$ and $\phi_{1_n}={\rm id}$. In addition, we
define the mapping $W: {\bf G} \to {\rm Sec}(\Sigma)$ whose values $W[G]$ are
arbitrary sections. However, we can impose supplemental conditions such that
the set of pairs $(W[G], \phi_G)$ should form a subgroup in ${\cal G}$.
Indeed, if we assume that
\begin{equation}\label{coco}
(W[G']\circ \phi_G)\times W[G]=W[G'G]\,, \quad W[1_n]={\rm Id}\,,
\end{equation}
then we find that the group multiplication,
\begin{equation}
(W[G'],\phi_{G'})*(W[G],\phi_G)=(W[G'G], \phi_{G'G}),
\end{equation}
defines the group ${\cal G}[{\bf G}]\subset {\cal G}$ of these pairs
which is in fact a representation of the group ${\bf G}$ {\em induced} by
${\bf G}_o$ since $W[G](\hat x)$ are elements of this group.

In the case of the matrices $(\ref{W})$ we identify
$W[G](\hat x)=W(G, \hat x)$ and from Eq. (\ref{compW}) it results that the
condition (\ref{coco}) is fulfilled. The conclusion is that the
transformations $(\ref{trx})$ and $(\ref{try})$ define an induced representation
in the sense outlined above.


\end{document}